\documentclass[prb,aps,twocolumn,floatfix,superscriptaddress]{revtex4}
\usepackage{graphicx}
\usepackage{subfigure}
\usepackage{bm}
\usepackage{amsmath}
\usepackage{tabularx}
\usepackage{array}
\usepackage{multirow}
\usepackage{float}
\usepackage{hyperref}
\begin{document}
\title{Self-protected nanoscale thermometry based on spin defects in silicon carbide }
\author{Yu Zhou}
\affiliation{Division of Physics and Applied Physics, School of Physical and Mathematical Sciences, Nanyang Technological University, Singapore 637371, Singapore}
\author{Junfeng Wang}
\affiliation{Division of Physics and Applied Physics, School of Physical and Mathematical Sciences, Nanyang Technological University, Singapore 637371, Singapore}
\author{Xiaoming Zhang}
\affiliation{Division of Physics and Applied Physics, School of Physical and Mathematical Sciences, Nanyang Technological University, Singapore 637371, Singapore}
\author{Ke Li}
\affiliation{Division of Physics and Applied Physics, School of Physical and Mathematical Sciences, Nanyang Technological University, Singapore 637371, Singapore}
\author{Jianming Cai}
\email[]{jianmingcai@hust.edu.cn}
\affiliation{School  of  Physics $\&$ Center  for  Quantum  Optical  Science,Huazhong University of Science and Technology,  Wuhan 430074,  P. R. China}
\author{Wei-bo Gao}
\email[]{wbgao@ntu.edu.sg}
\affiliation{Division of Physics and Applied Physics, School of Physical and Mathematical Sciences, Nanyang Technological University, Singapore 637371, Singapore}
\affiliation{MajuLab, CNRS-Universit\'{e} de Nice-NUS-NTU International Joint Research Unit UMI 3654, Singapore}
\affiliation{The Photonics Institute and Centre for Disruptive Photonic Technologies, Nanyang Technological University, 637371 Singapore, Singapore}
\begin{abstract}
Quantum sensors with solid state electron spins have attracted considerable interest due to their nanoscale spatial resolution. A critical requirement is to suppress the environment noise of the solid state spin sensor. Here we demonstrate a nanoscale thermometer based on silicon carbide (SiC) electron spins. We experimentally demonstrate that the performance of the spin sensor is robust against dephasing due to a self-protected mechanism from the intrinsic transverse electric field of the defect. The SiC thermometry may provide a promising platform for sensing in a noisy environment, e.g. biological system sensing. 
%, which is a superior character of SiC divacancies as compared to NV centers in diamond in this application.  
%term of the Hamiltonian of PL5 can suppress the effect of longitudinal magnetic field perturbation, and therefore enables a %self-protected mechanism to makes it robust to the noise. 
%High spatial diffraction limit resolution and mK sensitivity thermometry is critical to practical applications in the realm of %semiconductor industry and biological systems. Moreover, the self-protected mechanism helps to further improve the %sensitivity  which is valuable to quantum sensing and quantum metrology. 
\end{abstract}

\maketitle

Nanoscale thermometry has been demonstrated based on various of systems like quantum dot\cite{quantumdot,quantumdot2}, nanoparticle\cite{np,np2}, NV (nitrogen vacancy) center spin in diamond \cite{pnas,nanolett,livingcell,wang} due to its significant benefits to microelectronics and bio-application \cite{luminescence}\cite{atnanoscale}. Recently, electron spins in silicon carbide have been found to be optically addressable and show superior coherence properties \cite{polytype,roomtemperature_cohere}\cite{coherentcontrolsiv}. Besides the favorable features of both CMOS and bio-compatibility, sillicon carbide provides a large number of types defects that can be used as candidates for a spin sensor, including PL1-PL6 in 4H-SiC\cite{roomtemperature_cohere},QL1-QL6 in 6H-SiC\cite{polytype} and Ky5 in 3C-SiC\cite{3c}.  As compared with quantum dot and NV center, nanoscale and highly sensitive thermometry based on a semiconductor material silicon carbide maybe more fascinating because of its versatility in production and widely application in the realm of electronic devices\cite{sic}. Moreover, unlike NV center which has 4 possible orientations in bulk diamond\cite{nvreview}, one type of divacancy spins in silicon carbide has the same orientation which improves the sensitivity in varies of sensing application by using an ensemble of divacancy spins. 

One main challenge for quantum sensors is to improve the sensitivity of quantum metrology against environment noise. Several methods have been developed such as spin echo, dynamical decoupling \cite{dd1,dd2}\cite{cai}and quantum error correction\cite{quantumerror,quantumerror2}. These active methods usually make experiments more involved and suffer from certain limitations. In this Letter, we demonstrate high sensitivity temperature sensing based on the electron spins in 4H-SiC divacancies. Especially, the transverse electric field in such defects can suppress the effect of longitudinal magnetic field noise, leading to an improved sensitivity. The self-protected mechanism against decoherence provides an appealing route for scenarios where active methods for suppressing noise may not be suitable.

%Besides these active methods, a non zero transversal component of zero filed splitting E in silicon carbide divacancy %electron spins will induce a energy gap between S=1 spin system. This energy gap enables a self protection against the  %magnetic fluctuation noise and improves the thermometry sensitivity.

\textit{Theory} -- Here we are considering PL5  defect in SiC\cite{roomtemperature_cohere},  which is a basal C$_{1h}$ symmetry divacancy showing high optically detected magnetic resonance (ODMR) contrast at room temperature\cite{roomtemperature_cohere}.  The ground state shows a spin-1 character with the basis written as $\{ |\uparrow\rangle,|0\rangle,|\downarrow\rangle \} $. The three ground states split at zero magnetic field, resulting in two ODMR resonance spectrum at $D+E$ and $D-E$. The spin Hamiltonian can be written as
\begin{equation}
H=H_{0}+H',
\end{equation}
where
\begin{equation}
H_{0}=\hbar D(T) S_{z}^{2}+\hbar E_{x}(S_{x}^{2}-S_{y}^{2});
\end{equation}
\begin{equation}
H'=g\mu_{B}B_{z}S_{z}+d_{z}\Pi_{z}S_{z}^{2},
\end{equation}
For $H_{0}$ term, $D(T)$ is the temperature-dependence zero field splitting, $E_{x}$ is the transverse electric field due to the lower symmetric of basal defect \cite{roomtemperature_cohere}, and $S_{x,y,z}$ represents the electronic spin operator. %For simplicity, we define the direction of the electric field as the $\hat{x}$ axis. 
The eigenstates of $H_{0}$ is $|\pm\rangle=\frac{1}{\sqrt{2}}(| \uparrow\rangle \pm|\downarrow\rangle)$ , and $|0\rangle$, with corresponding eigenvalues $D\pm E_x,0$. For $H'$ term, $g= 2.00$ is the electron g factor, $\mu_{B}$ is the Bohr magneton, $B_{z}$ represents the external magnetic field fluctuation. The effect of the transverse component of an external magnetic fluctuation is suppressed by the zero field splitting.  Thus, the magnetic field fluctuation can be described by the longitudinal component $B_{z}$ \cite{timekeeping}. In addition, we also consider the external electric field fluctuation as denoted as $\Pi_{z}$. 

We first perform ODMR measurement by applying an external microwave with frequency $\omega$ that is nearly on resonant with the transition $|0\rangle \longleftrightarrow |+\rangle$. In the rotating frame, the total Hamiltonian under rotating-wave approximation can be rewritten as \cite{sm}:
\begin{equation}
H_{rot}=\hbar\begin{bmatrix}
\Delta+\Pi_{z}'-E_{x}+B'_{z} &\frac{1}{\sqrt{2}}\Omega & E_{x} \\\frac{1}{\sqrt{2}}\Omega &0 & \frac{1}{\sqrt{2}}\Omega \\ E_{x} &\frac{1}{\sqrt{2}}\Omega & \Delta+\Pi_{z}'-E_{x}-B'_{z},
\end{bmatrix}
\label{eq:HAM}
\end{equation}
where the detuning is $\Delta=D(T)+E_{x}-\omega$, $B_{z}'=\frac{g\mu_{B}}{h}B_{z}$, $\Pi_{z}'=\frac{d_{z}\Pi_{z}}{h}$. One of the most remarkable features of the system Hamiltonian Eq.(\ref{eq:HAM}) is the off-diagonal term $E_{x}$, which induces the transition between $|{\uparrow}\rangle$ and $|{\downarrow}\rangle$. It would thus average out the effect of the magnetic field fluctuation acting on $\vert {\uparrow}\rangle, \vert {\downarrow}\rangle$, which essentially plays the role of continuous dynamical decoupling.

%We will discuss later how this term can protect the system from longitudinal magnetic field perturbation.

To perform measurement of temperature, we adopt a Ramsey scheme. We first initialize the spin into the state $|0\rangle$, and apply a $\pi/2$ pulse  to drive the system into the superposition sate of $|0\rangle$ and $|+\rangle$ with equal amplitude: $U_{\pi/2}|0\rangle=\frac{1}{\sqrt{2}}(|0\rangle-i|+\rangle)$. After a free evolution for time $\tau$, we appy another $\pi/2$ pulse to map the phase information to the state population. It can be shown that the final population of $|0\rangle$ state is as follows \cite{sm}
\begin{eqnarray}
P_{0}'=\frac{1}{2}\left[ 1-\cos2\pi\left (\Delta+\Pi_{z}'+\frac{B_{z}'^{2}}{2E_{x}}\right )\tau\right].
\end{eqnarray}
The oscillation of $P_{0}'$ is dominated by the detunning $\Delta=D(T)+E_{x}-\omega$. Since $D(T)$ changes with temperature, while the noise would mainly reduce the signal contrast and has negligible effect on the oscillation frequency. Therefore, the oscillation frequency of $P_{0}'$ can be used to determine the value of temperature.

The sensitivity for the measurement of temperature is determined by the coherence time of the spin sensor. The main noise that would affect the coherence time in the present experiment is the longitudinal external magnetic field fluctuation.  In the free evolution step, the unexpected $B_{z}$ has opposite effects on spin ${\uparrow}$ and ${\downarrow}$ by giving a positive or negative phase for $|{\uparrow}\rangle$ or $|{\downarrow}\rangle$ state. As can be seen from the Hamiltonian in Eq.(\ref{eq:HAM}), one of the effects of transverse electric field is to flip the $|{\uparrow}\rangle$ and $|{\downarrow}\rangle$ continuously. Therefore, the direction of the magnetic field experienced by the electron spin is changing continuously. If the electric field is large enough, the average effect of $B_{z}$ vanishes. The effect induced by the environment noise $B_{z}$ is thus reduced to the order of $O(\frac{B_{z}^{'2}}{E_{x}})$ as compared with $O(B'_{z})$ when $E_{x}$ is negligible \cite{pnas}. This provide a self-protected mechanism even without requiring extra operation as conventional active dynamical decoupling schemes.

\textit{Experiment: Sample characterization} -- First, we describe our experimental setup. Here we use a 950nm laser focusing on the sample through an infrared objective with N.A. of 0.8. Fluorescence above 1000nm is collected and guided through a multimode fiber to an infrared photon diode. To manipulate the spin state, the microwave with tunable frequency is fed to the sample through a microwave antenna fabricated on the sample. The antenna (Au/Pt) has a ring shape structure with 80um inner diameter. To reduce the microwave noise, we modulate the microwave with frequency 20Hz using switch and the same signal is used for lock-in detection of the fluorescence from the defects.

 \begin{figure}
	\centering
	\includegraphics[height=6cm]{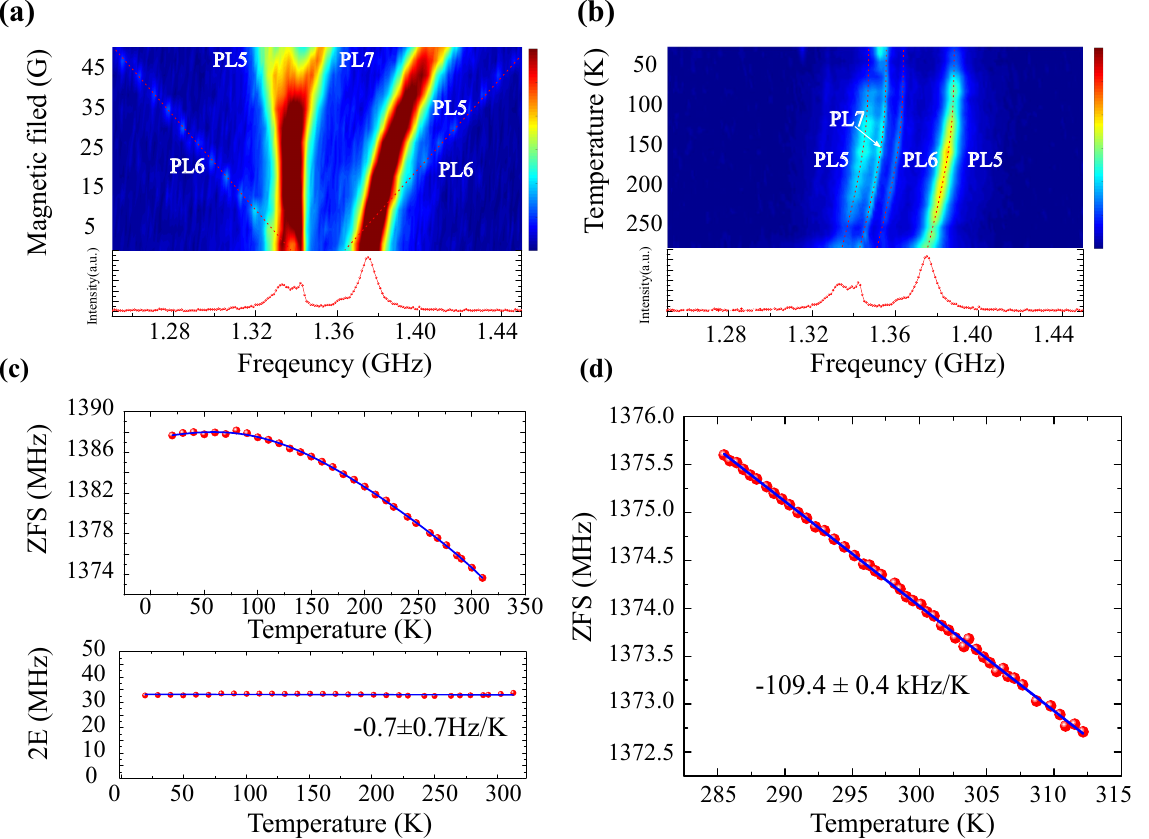}
	\caption{(a) Room temperature ODMR spectrum map of 4H-SiC divacancy spins PL5 with respect to B field along $c$ axis. Lower panel shows a line cut at zero magnetic field. PL5, PL6, PL7 represent different defects types in SiC. (b) ODMR spectrum map as a function of temperature. Dashed line is drawn as a guide of eye. Lower panel shows a line cut at 300K. (c) Upper panel: ZFS (Zero Filed Splitting) parameter shift for PL5 right branch ODMR spectrum as a function of temperature. Lower panel: Transverse components of the PL5 ZPL with respect to the temperature, which shows that it stays roughly in a constant level. (d) ZFS shift at near room temperature. A linear shift is used for the fitting. }\label{fig:odmr}
\end{figure}

With the setup described, we first characterize the ODMR frequency of the divacancy defects. As shown in Fig. \ref{fig:odmr}(a) lower panel, different microwave frequency peaks are observed in the ODMR spectrum at zero magnetic field. To further identify these peaks, we measured the ODMR spectrum as a function of the applied magnetic field in the $z$ direction, as shown in  Fig. \ref{fig:odmr}(a) upper panel.  Due to different symmetry property of the defects, these ODMR resonances show different diverging behavior. For example, at zero magnetic field, two resonance transitions of $c$ axis divacancy have a small splitting. When there is a static magnetic field along the $c$ axis, two transitions split at the slope of 2.8MHz/G as marked using dash line in the figure. This shows the same behavior as NV center \cite{nvreview}. For these transitions, we identify them as PL6.  While for $C_{1h}$ symmetry PL5, two resonance transitions will bend and the left branch of the PL5 resonance is mixed with PL7 and the right branch is alone. All thermometry experiment in this paper is based on the right branch resonant transition of PL5 ($|0\rangle  \longleftrightarrow |+\rangle$) because it is well isolated with other transitions.

Next, we proceed to measure temperature dependence of the ODMR spectrum. Due to thermal expansion and electron-phonon interactions, transition resonance frequencies tend to increase when the temperature decreases \cite{prb_shift}. In the ODMR spectrum scan shown in Fig. \ref{fig:odmr}(b), all transitions belonging to PL5-PL7 exhibit similar behavior, which shows that such resonance frequency shift is their intrinsic property. The temperature dependence of the PL5 Zero Field Splitting (ZFS) parameter $D$ ranging from 20K to 300K is shown in Fig.1(c) and the nonlinear shift is fitted with a fifth-order polynomial. The transverse electric field $E_x$ is also estimated from the difference between the left branch ($D-E_x$) and right branch ($D+E_x$). By linear fitting with a slope of -0.7Hz/K, it indicates that the transversal component $2E_x$ doesn't have an obvious shifting, and the shifting of the ZFS is mainly due to the axial component $D$ which shows the same behavior as NV center in diamond \cite{tempereaturedependence}. In oder to get more detailed shifting of the $D$ value with respect to the temperature, we measure the ODMR resonance of PL5 right branch ($D+E$) near room temperature (Fig. \ref{fig:odmr}(d)). It follows a good linear relationship with respect to the temperature at a slope of $dD/dT = -109.4 \pm 0.4 \mbox{kHz}/\mbox{K}$. It indicates that the OMDR transition in PL5 is more sensitive to temperature, as compared to what has been measured using NV center in diamond ($dD/dT \sim -74.2\mbox{kHz}/\mbox{K}$) \cite{tempereaturedependence}.

 \begin{figure}
	\centering
	\includegraphics[height=6cm]{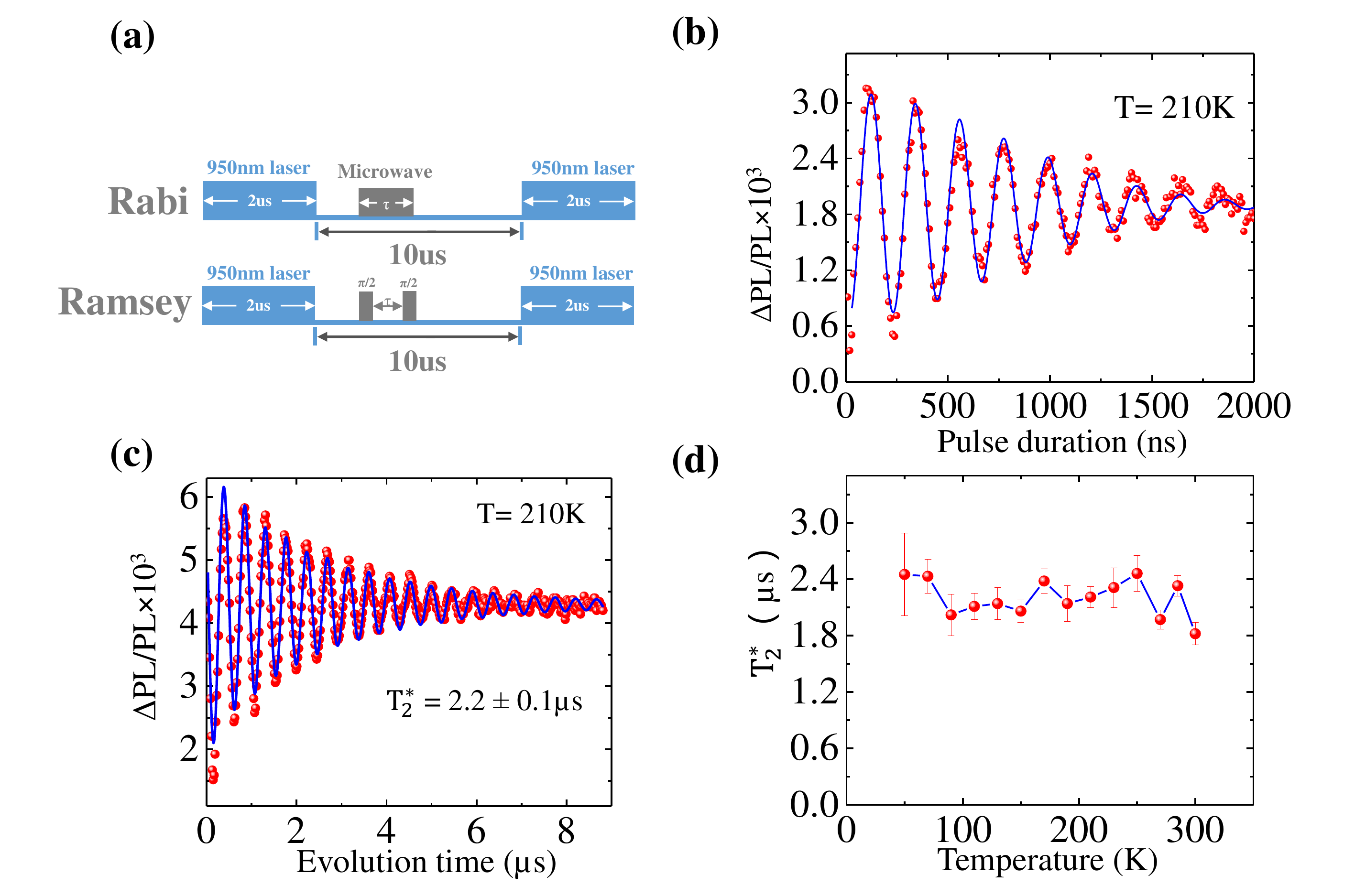}
	\caption{(a) Rabi and Ramsey pulse sequence. A 950nm laser pulse is used for spin initialization and spin readout. One microwave pulse with variable pulse length is used for Rabi oscillation measurement. Two $\frac{\pi}{2}$-pulses with a variable delay are used for Ramsey fringes. (b) Rabi oscillation of PL5. Fluorescence change has been recorded as a function of microwave pulse length. A decayed sinuous function is used to fit the experimental data. The measurement is performed at temperature 210K.  (c) Ramsey fringe. The microwave frequency has a detuning of 2MHz with respect to the resonance frequency. The oscillations shows decayed sinuous oscillation with a period of $500ns$. The decay time shows that the dephasing time T$_2^*$ is $2.2\mu s$. (d) Inhomogeneous dephasing time with respect to temperature.}
\end{figure}

\textit{Coherent control} -- The frequency shift of ZFS shown above indicates that SiC divacancies can be potentially used as a temperature sensor by using Ramsey interferometry.  As a first step, we demonstrate the coherent control of electron spins (Fig. 2).  The pulse sequence for Rabi oscillation and Ramsey fringe is shown in Fig.2(a).  First, a 2 $\mu s$ $\frac{\pi}{2}$-pulse is used to initialize the spin. Following the initialization pulse, the 20 Hz modulated Rabi or Ramsey microwave pulse is applied to coherently manipulate the spin state. Finally the spin state is read out by another 2 $\mu s$ $\frac{\pi}{2}$-pulse. The population change of the  final state will result in a change of PL intensity $\Delta PL$, which is detected with lock-in method. Fig.2(b) and (c) shows Rabi oscillations and Ramsey fringes respectively when the temperature was 210K. The Ramsey fringes is fitted with the equation 
\begin{equation}
I=a\exp\left[-\left(\frac{t}{T_2^*}\right)^{n}\right]\cos(2\pi ft+\varphi)+b
\end{equation}
where $a$, $n$, $\varphi$ and $b$ are free parameters, $t$ is the evolution time and $T_2^*$ is dephasing time. The oscillation of Ramsey measurement was induced by a microwave with a detuning of 2MHz from resonance. Inhomogeneous spin-dephasing time is a critical factor in the dc-strain sensing and temperature sensing\cite{electronic_pro,pnas}.  For each temperature, we measured the Ramsey fringes and extracted the inhomogeneous spin-dephasing time T$_2^*$ as shown in Fig.2(d). The result shows that T$_2^*$ remains almost constant at around 2$\mu$s at the temperature ranging from 4K to 300K. 

  \begin{figure}
 	\centering
 	\includegraphics[height=3.8cm]{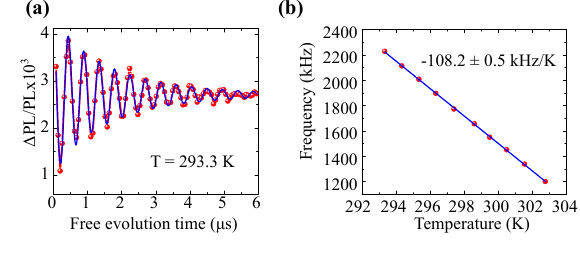}
 	\caption{(a)Ramsey fringes at 293.3K. Oscillation is fitted with Eq.(6). %Inhomogeneous dephasing time was 1.8$\mu$s.  
 	(b)Ramsey oscillation frequency $\Delta (T)=D(T)+E_{x}-\omega$ as a function of temperature when the microwave frequency $\omega$ is fixed. $\Delta (T)$ is fitted with a linear function with a slope of $-108.2\pm0.5\mbox{kHz}/\mbox{K}$. }
 \end{figure}

Next we demonstrate thermometry based on Ramsey fringes method. The frequency of Ramsey fringes is induced by detuning $\Delta$.  As explained above, the temperature shift will result in a linear change of D value as shown in Fig.2(d) which is directly related to the oscillations frequency of the Ramsey fringes. As shown in the Fig.3(a), Ramsey measurement of PL5 shows good oscillations without strong coupling with other unwanted nuclear spins. The data is fitted with equation (6).  Fig.3(b) displays the Ramsey oscillation frequency follows a good liner relationship with respect to the temperature change. The slope $-108.2\pm0.5\mbox{kHz}/\mbox{K}$ also matches with the measured temperature dependence shift $-109.4\pm0.4\mbox{kHz}/\mbox{K}$ of the $D$ value obtained from the ODMR resonance spectrum. The Ramsey based method can be applied to a wide range of temperature, as shown in Fig. 1(c). The thermal sensitivity $\eta$ has a estimated value 205.6mK/Hz\textsuperscript{1/2} following the method as described in previous references \cite{pnas}. 

% We estimate sensitivity of the Ramsey based thermometry by the following
% equation\cite{pnas}
% \begin{equation} 
% \eta=\sqrt{\frac{2(p_{0}+p_{1})}{(p_{0}-p_{1})^{2}}}\thinspace\frac{1}{2\pi\thinspace\frac{dD}{dT}\exp\thinspace(-(\frac{t}{T_2^*})^{n})\thinspace\sqrt{t}}\label{eq: thermal sensitivity}
% \end{equation}
% where $p_{0}$ and $p_{1}$ are the photon counts per measurement
% shot for the bright and dark spin states, respectively. n is the fitting parameter in equation (6). In the experiments,
% we used the infrared objective (NA = 0.8) and the obtained $p_{0}$ and
% $p_{1}$ values were about 99.4 and 99.16, respectively. Thus derived
% thermal sensitivity $\eta$ of the Ramsey was 205.6mK/Hz\textsuperscript{1/2}. 

% However, the Ramsey based thermometry limits the coherence to the inhomogeneous spin lifetime T$_2$$^ *$ which is 1.8$\mu$s at room temperature. Another pulse sequence pi/2-$\tau$-2pi-$\tau$-pi/2was first proposed as a timekeeping sequence in NV center in diamond \cite{timekeeping}. Thermometry based on this TE sequence was also performed in NV center\cite{pnas} and in bio-sensing\cite{livingcell}. As an echo based technique, this sequence is robust to the spin and magnetic environment, the corresponding sensitivity 188.5mK/Hz\textsuperscript{1/2}. Moreover, the coherence would be longer than inhomogeneous spin-dephasing time which enables a better sensitivity.

\textit{Self-protection effect} -- As described in the previous sections, the defect types we are using have a large transverse electric field $E_x$, which will protect the temperature sensor against the environment magnetic field noise. We experimentally verify such a self-protection mechanism as provided by $E_{x}$. In the experiment, we vary the magnetic field around the sample with a randomly changed value between $-B$ to $B$ (where $B$ is the maximum amplitude of the magnetic field fluctuation) at a frequency of 100Hz, and then observe how the Ramsey oscillation decay changes, which represent the effect of the magnetic field fluctuation on T$_2^*$. 

The experiment results are shown in Fig.4. For SiC PL5, $E$ is measured to be $16.5$MHz as measured in Fig.1(c). Our theoretical simulation as shown in red dots (Fig.4) matches the experimental data quite well. The detailed simulation method and result is shown in supplementary material. When $E_x=0$, T$_2^*$ decreases dramatically for small amplitude as shown in the simulation. We note that this is in agreement with the previous experimental results based on NV center in diamond, where T$_2^*$ is found to be very sensitive to the magnetic field perturbation \cite{prbex}. Both the simulation and experimental data support our observation that the intrinsic non-zero electric field $E$ enables a self-protected mechanism for the electron spin coherence against magnetic field noise and therefore sustain a high sensitivity of the present temperature sensing.

 \begin{figure}
	\centering
	\includegraphics[height=6cm]{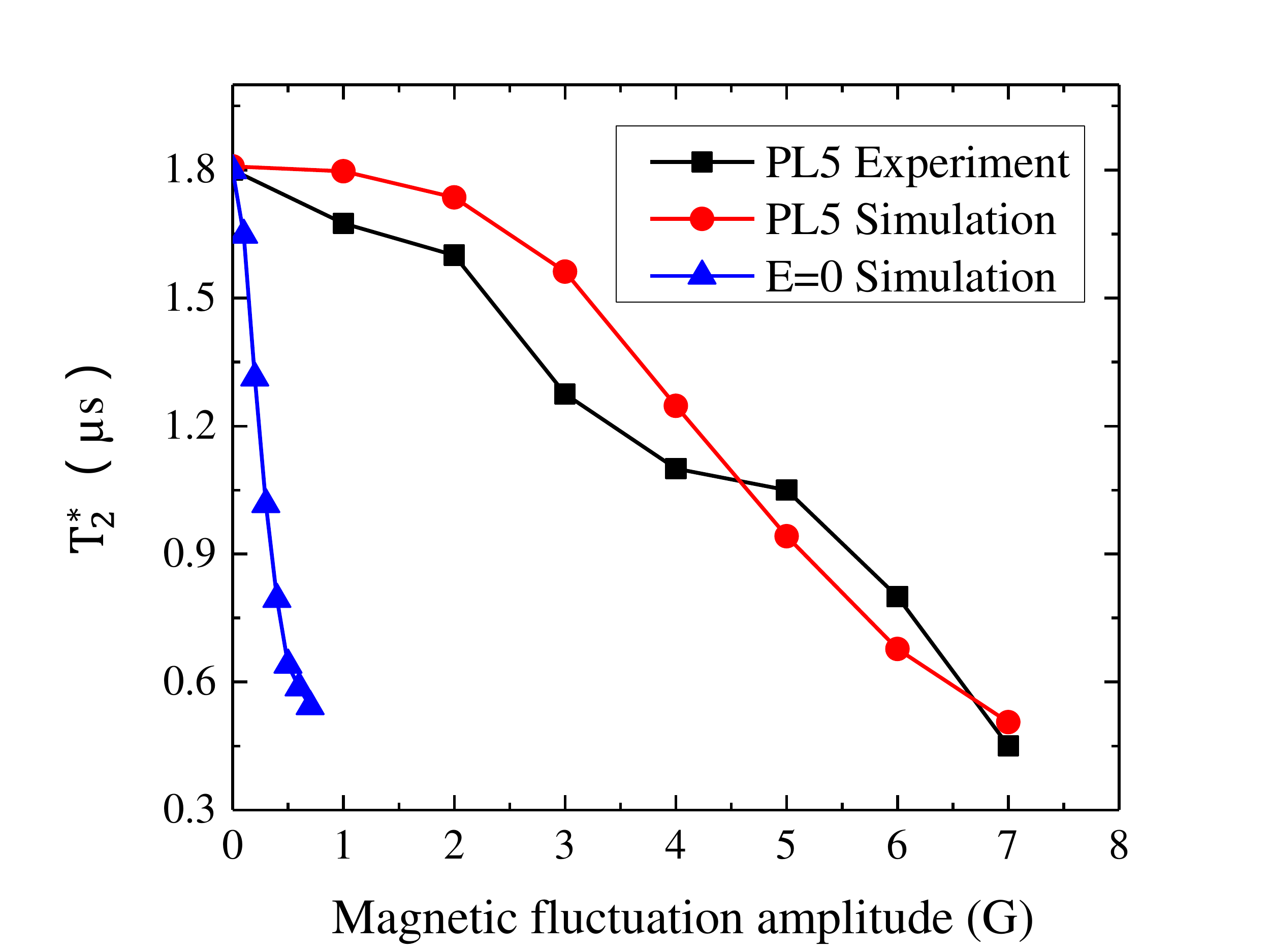}
	\caption{The relationship between T$_2^*$  and magnetic fluctuation amplitude (black dot for experiment data; red and blue dot for simulation). For PL5 defect with $E=16.5MHz$, T$_2^*$ decays slower than the case when $E=0$, which shows the self-protect effect by the large $E_{x}$ in SiC PL5 defects. }
\end{figure}

\textit{Conclusion} -- In summary, we have demonstrated SiC thermometry based on the coherence of electron spins with Ramsey pulse sequence.  Especially, they are protected by the intrinsic large transverse electric field of the defects, which simplifies the procedure for enhancing the sensor sensibility. Considering that the material itself is bio-compatible and also the required infrared laser beam would cause less damage to the bio-system\cite{irlaser}. This technique can be applied to the thermometry of a living cell if silicon carbide nano particles is used as in the nano-diamond case \cite{livingcell}. 

We acknowledges the support from the Singapore National Research Foundation through a Singapore 2015 NRF fellowship grant (NRF-NRFF2015-03) and its Competitive Research Programme (CRP Award No. NRF-CRP14-2014-02), Astar QTE project and a start-up grant (M4081441) from Nanyang Technological University. J.-M.C is supported by the National Natural Science Foundation of China (Grant No.11574103). Y.Z.,J.W.and X.Z. contributed equally to this work.

\section* {APPENDIX}

\section{Derivation of Eq.(4) in the main text}
In basis of $\{\uparrow,0,\downarrow\}$, the initial Hamiltonian $H=H_{0}+H'$ can be written in the matrix form:
\begin{equation} \tag{S1}
H=h\begin{bmatrix} 
D+\Pi_{z}'+B_{z}' &0 & E_{x} \\ 0 &0 & 0 \\ E_{x} & 0 & D+\Pi_{z}'-B_{z}'
\end{bmatrix}
\end{equation}
An external microwave with frequency $\omega$ that is nearly on resonant with transition $|0\rangle \longleftrightarrow |+\rangle$ is applied to the system. Then, we change to the new basis of $\{ +,0,- \}$, where $|\pm\rangle =1/\sqrt{2}(|0\rangle+|1\rangle)$, the total Hamiltonian adding the microwave term can be rewritten as
\begin{equation} \tag{S2}
H_{dr,\pm}=h\begin{bmatrix} 
D+\Pi_{z}'+E_{x} &\Omega\cos \omega t & B'_{z} \\ \Omega\cos  \omega t &0 & 0 \\ B'_{z} & 0 & D+\Pi_{z}'-E_{x}
\end{bmatrix}
\end{equation}
where $\Omega$ is the Rabi frequency, and $B'_{z}=\frac{g\mu_{B}}{h}B_{z}$, $\Pi_{z}'=\frac{d_{z}}{h}\Pi_{z}'$. As can be seen, the microwave would only drive the transition between $|0\rangle$ and $|+\rangle$, while $|-\rangle$ is decoupled. Then, we change to rotating frame \cite{timekeeping} with the operator $V=\exp{( i\omega t S_{z}^{2})}$, and define the detuning $\Delta=D+E_x-\omega $ with the assumption that $\Delta \ll D+E_x$. The new Hamiltonian under rotating wave approximation can be written as
\begin{equation} \tag{S3}
H_{rot,\pm}=h\begin{bmatrix}
\Delta+\Pi_{z}' &\Omega & B'_{z} \\ \Omega &0 & 0 \\ B'_{z} & 0 & \Delta+\Pi_{z}'-2E_{x}
\end{bmatrix}\\
\end{equation}
Going back to the basis $\{\uparrow,0,\downarrow\}$, the total Hamiltonian becomes
\begin{equation} \tag{S4}
H_{rot}=h\begin{bmatrix}
\Delta+\Pi_{z}'-E_{x}+B'_{z} &\frac{1}{\sqrt{2}}\Omega & E_{x} \\\frac{1}{\sqrt{2}}\Omega &0 & \frac{1}{\sqrt{2}}\Omega \\ E_{x} &\frac{1}{\sqrt{2}}\Omega & \Delta+\Pi_{z}'-E_{x}-B'_{z}
\end{bmatrix},
\end{equation}
which is Eq.(4) in the main manuscript. 

\section{Derivation of Eq.(5)}

The microwave pulse can be switched on or off by choosing Rabi frequency $\Omega$ or $0$. When the microwave is off, the Hamiltonian is:
\begin{equation} \tag{S5}
H_{rot}(0)=h\begin{bmatrix}
\Delta+\Pi_{z}'-E_{x}+B'_{z} &0 & E_{x} \\ 0 &0 & 0 \\ E_{x} & 0 & \Delta+\Pi_{z}'-E_{x}-B'_{z}
\end{bmatrix}.
\end{equation}	  
We recall that $\Omega\gg \Delta,B'_{z},\Pi_{z}'$, and the pulse lengths are much shorter than the free evolution time. Therefore, the terms including $\Pi_{z}', B_{z}', \Delta$ can be neglected during the pulse operations. When the microwave is on, the Hamiltonian in basis $\{\uparrow,0,\downarrow\}$ can be written as:
\begin{equation} \tag{S6}
H_{rot}(\Omega)=h\begin{bmatrix} 
-E_{x} &\frac{1}{\sqrt{2}}\Omega & E_{x} \\\frac{1}{\sqrt{2}}\Omega &0 & \frac{1}{\sqrt{2}}\Omega \\ E_{x} & \frac{1}{\sqrt{2}}\Omega &-E_{x}
\end{bmatrix}.
\end{equation}
For an initial state $|0\rangle$, we first apply a $\pi/2$ pulse $U_{\pi/2}=e^{-i\frac{\pi}{4 \Omega}H_{rot}(\Omega)/\hbar}$ that drives the system to superposition sate of $|0\rangle$ and $|+\rangle$:
\begin{equation} \tag{S7}
U_{\pi/2}|0\rangle=\frac{1}{\sqrt{2}}(|0\rangle-i|+\rangle).
\end{equation}
Then, the system undergoes a free evolution for time $\tau$ represented by the operator $U_{\tau}=e^{-iH_{0,rot}(0)\tau/\hbar}$. The state becomes:
\begin{small}
	\begin{equation}
	\begin{matrix}
	U_{\tau} U_{\pi/2}|0\rangle=\frac{1}{\sqrt{2}}|0\rangle 
	+ \frac{e^{-2\pi i(\Delta+\Pi_{z}'-E_{x})\tau}}{2i}\\\ (\cos 2\pi\delta\tau-\frac{E_{x}+B'_{z}}{\delta}i\sin2\pi\delta\tau)|\uparrow\rangle
\\	+(\cos 2\pi\delta\tau-\frac{E_{x}-B'_{z}}{\delta}i\sin2\pi\delta\tau)
	|\downarrow\rangle,
	\end{matrix}
	\end{equation}
\end{small}
where $\delta=\sqrt{E_{x}^{2}+B_{z}^{'2}}\approx E_{x}(1+\frac{B_{z}^{'2}}{2E_{x}^{2}})$. Finally, we apply another $\pi/2$ pulse before measurement of the spin state population. The final amplitude of the spin state $|0\rangle$ can be calculated by $A_{0,final}=\frac{1}{2}-\frac{1}{2}i(A_{\uparrow}'+A_{\downarrow}')$, where $A_{\uparrow}'$ ($A_{\downarrow}'$) is the amplitude of $|{\uparrow}\rangle$ ( $|{\downarrow}\rangle$) before the final $\pi/2$ rotation. If one neglect terms higher than $O(B_{z}/E_{x})$, it becomes:
\begin{equation} \tag{S9}
A_{0,final}=\frac{1}{2}(1-ie^{-2\pi i (\Delta+\Pi_{z}'+\frac{B_{z}^{2}}{2E_{x}})\tau}),
\end{equation}

and the value of final population of $|0\rangle$ state is:
\begin{equation} \tag{S10}
P_{0,final}'=\frac{1}{2}-\frac{1}{2}\cos2\pi (\Delta+\Pi_{z}'+\frac{B_{z}^{2}}{2E_{x}}) \tau,
%+\frac{1}{2}(\Im+\frac{B_{z}^{'2}}{2E_{x}})\tau\sin{\Delta\tau}.
\end{equation}
which is Eq.(5) in the main manuscript.

\section{Numerical simulation}
\subsection{Ramsey fringes \textit{vs} $E_{x}$}
In order to numerically verify the role of $E_{x}$ in the sensing dynamic, we simulate the Ramsey fringes for different values of $E_{x}$. Fig.2 shows the average of $P_{0,final}$ over 1000 runs versus free evolution time $\tau$ for different value of $E_{x}$. For each run, $B_{z}'$ is randomly chosen and kept as constant during the evolution.  %and $\Pi_{z}'=0$ is set to be $0$. % 
In our simulation, $B_{z}'$ is assumed to follow the standard normal distribution, and the standard deviation is set as $\sigma_{B_{z}'}=0.2$MHz. As can be seen, the effect of the magnetic field noise $B_{z}'$ is gradually weeken when the value of $E_{x}$ increase. For $E_{x}=16.5$MHz, the magnetic field noise already have no observable effect on the evolution. 

\begin{figure}
	\centering
	\includegraphics[height=6cm]{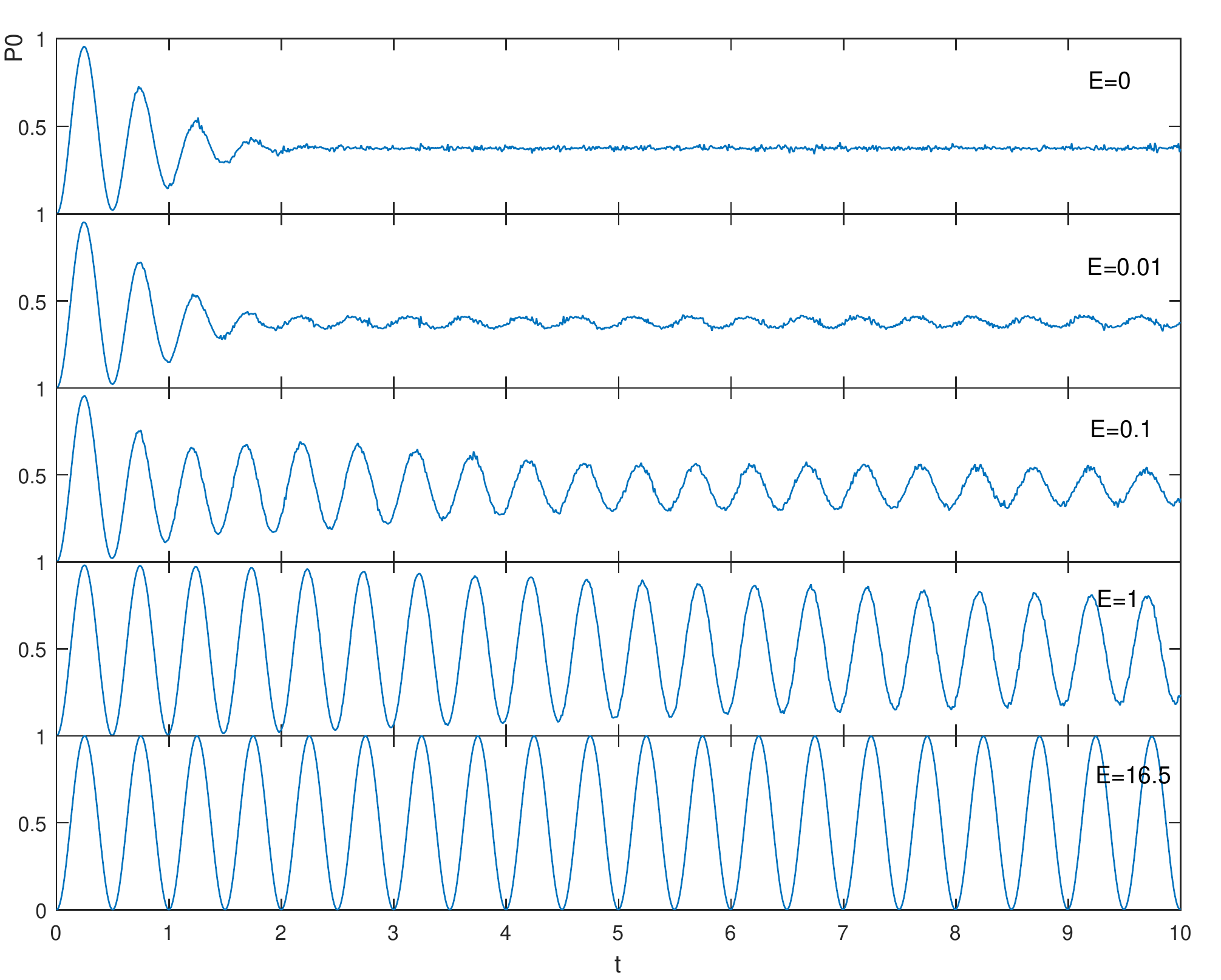}
	\caption{Simulation of Ramsey fringes for different values of $E_{x}$ with unit of MHz. The other parameters are set to be $\Delta=2$MHz, $ \Pi_{z}'=0$MHz. For each value of $E_{x}$, $B_{z}'$ is chosen 1000 times randomly from a standard normal distribution of $\sigma=0.2$MHz.}
\end{figure}

\subsection{$T_{2}^{\star}$ \textit{vs} AC Magnetic Field }

In the main text Figure. 4, both experimental and numerical results of  $T_{2}^{\star}$ for different applied AC magnetic fields have been shown. The simulation is done with Hamiltonian Eq.(S1) following the Ramsey pulse sequence discussed in the main manuscript. In our experiment, the magnetic field can only applied perpendicular to the sample. Since the $c$-axis of PL5 is at $109.5^{\circ}$ from the normal of the sample\cite{strain}, $B_{z}$ in Eq.(S10) should be $B_{z}'=\cos(109.5^{\circ})B_{applied}$, where $B_{applied}$ is total amplitude of the magnetic field applied by an electronic magnet. There are two random parameters in Eq.(S10), namely $\Pi_{z}'$ and $B_{z}'$. For $\Pi_{z}'$, we assume it follows the standard normal distribution with zeros average value. The standard deviation $\sigma_{\Pi_{z}'}$ is determined by the experimental result $T_{2,B_{z}=0}^{\star}=1.8\mu s$ when $B_{applied}=0$MHz. When the magnetic field fluctuation is negligible, one has the relation\cite{prbex}:$    T_{2,B=0}^{\star}=\frac{1}{\sqrt{2}\pi\sigma_{\Pi_{z}'}}$. For $B_{z}'$, we assume it to be white noise between the maximum ($+B$) and minimum ($-B$) allowed value. For each value of $B$, we average the evolution of 1000 times of run for different randomly chosen $\Pi_{z}'$ and $B_{z}'$. The value of $T_{2}^{\star}$ is obtained by the fitting of the average evolution. 

\section{Thermo Echo measurement}
%  \begin{figure}
% 	\centering
% 	\includegraphics[height=3.5cm]{figure4_1.pdf}
% 	\caption{(a)TE(Temperature Echo) sequence at room temperature. The oscillation was induced by detuning of the microwave frequency for 2MHz. (b)TE sequence oscillation frequency respect to temperature.}
% \end{figure}

\begin{figure}
	\centering
	\includegraphics[height=4cm]{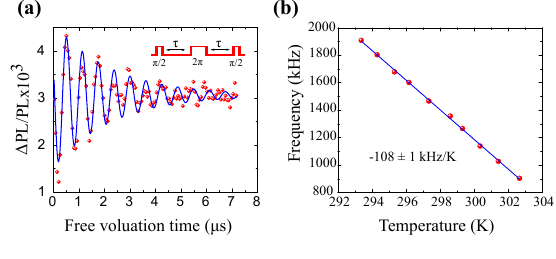}
	\caption{(a)TE (Temperature Echo) sequence at room temperature. (b) TE sequence oscillation frequency respect to temperature.}
\end{figure}

One of the most widely used dynamically decoupled schemes for thermometry is the Thermo Echo (TE) sequence \cite{pnas,timekeeping}. TE is the Ramsey pulse sequence adding a $2\pi$ pulse in the middle of free evolution. With this extra $2\pi$ pulse, states $|{\uparrow}\rangle$ and $|{\downarrow}\rangle$ exchange their population, making the asymmetric phase accumulation of $B'_{z}S_{z}$ term before and after the $2\pi$ rotation cancel each other. For NV center in diamond, TE can increase the $T_{2}^{\star}$ by at least an order of magnitude as compare to Ramsey measurement. In our system of PL5 divacancy in SiC, however, the $T_{2}^{\star}$ is not expected to increase with TE, because the effect of $B_{z}'$ has already been suppressed for Ramsey measurement, which servers as another evidence of a self-protection mechanism.

We experimentally test the TE pulse sequence for thermometry. As shown in the Fig.2(a), the extracted coherence time is 2.3$\mu$s which is in the same order of the dephasing time for Ramsey sequence. This agrees well with our theoretical model. The frequency follows a good linear relationship with the temperature change which is in the Fig.2(b). The slope -108$ \pm$1kHz/K again matches with the temperature dependence of the D value very well. %However, the coherence time of both Ramsey and TE sequence are still not comparable to $T_{2}=39\mu s$   measured in \cite{coherence} . This is because the sample is as grown without any isotope purification, which can induce the significant electrical noise.

\end{document}